\documentclass[11pt]{article}
\usepackage{moriond,epsfig}

\begin{document}
\vspace*{4cm}
\title{SEARCHES FOR LEPTON FLAVOUR AND LEPTON NUMBER VIOLATION\\
IN KAON DECAYS}

\author{Evgueni Goudzovski}

\address{Centre for Cosmology, Particle Physics and Phenomenology,\\
Catholic University of Louvain, B-1348 Louvain-la-Neuve, Belgium\\
and\\School of Physics and Astronomy, University of Birmingham,\\
Edgbaston, Birmingham, B15 2TT, United Kingdom}

\maketitle\abstracts{Searches for lepton flavour and lepton number
violation in kaon decays by the NA48/2 and NA62 experiments at CERN
are presented. A new measurement of the helicity suppressed ratio of
charged kaon leptonic decay rates $R_K=\Gamma(K_{e2})/\Gamma(K_{\mu
2})$ to sub-percent relative precision is discussed. An improved
upper limit on the lepton number violating
$K^\pm\to\pi^\mp\mu^\pm\mu^\pm$ decay rate is also presented.}

\section{Introduction}

In the Standard Model (SM) the decays of pseudoscalar mesons to
light leptons are helicity suppressed. In particular, the SM width
of $P^\pm\to\ell^\pm\nu$ decays with $P=\pi,K,D,B$ (denoted $P_{\ell
2}$ in the following) is
\begin{equation}
\Gamma^\mathrm{SM}(P^\pm\to\ell^\pm\nu) = \frac{G_F^2 M_P
M_\ell^2}{8\pi} \left(1-\frac{M_\ell^2}{M_P^2}\right)^2
f_P^2|V_{qq\prime}|^2, \label{eq:sm}
\end{equation}
where $G_F$ is the Fermi constant, $M_P$ and $M_\ell$ are meson and
lepton masses, $f_P$ is the decay constant, and $V_{qq\prime}$ is
the corresponding Cabibbo-Kobayashi-Maskawa matrix element. Although
the SM predictions for the $P_{\ell 2}$ decay rates are limited by
hadronic uncertainties, their specific ratios do not depend on $f_P$
and can be computed very precisely. In particular, the SM prediction
for the ratio $R_K=\Gamma(K_{e2})/\Gamma(K_{\mu 2})$ of kaon
leptonic decay widths inclusive of internal bremsstrahlung (IB)
radiation is~\cite{ci07}
\begin{equation}
\label{Rdef} R_K^\mathrm{SM} = \left(\frac{M_e}{M_\mu}\right)^2
\left(\frac{M_K^2-M_e^2}{M_K^2-M_\mu^2}\right)^2 (1 + \delta
R_{\mathrm{QED}})= (2.477 \pm 0.001)\times 10^{-5},
\end{equation}
where $\delta R_{\mathrm{QED}}=(-3.79\pm0.04)\%$ is an
electromagnetic correction due to the IB and structure-dependent
effects.

Within certain two Higgs doublet models (2HDM of type II), including
the minimal supersymmetric model (MSSM), $R_K$ is sensitive to
lepton flavour violating (LFV) effects appearing at the one-loop
level via the charged Higgs boson ($H^\pm$)
exchange~\cite{ma06,ma08}, representing a unique probe into mixing
in the right-handed slepton sector~\cite{el09}. The dominant
contribution due to the LFV coupling of the $H^\pm$ is
\begin{equation}
R_K^\mathrm{LFV}\simeq
R_K^\mathrm{SM}\left[1+\left(\frac{M_K}{M_H}\right)^4
\left(\frac{M_\tau}{M_e}\right)^2 |\Delta
_R^{31}|^2\tan^6\beta\right], \label{rk_lfv}
\end{equation}
where $\tan\beta$ is the ratio of the two Higgs vacuum expectation
values, and $|\Delta_{R}^{31}|$ is the mixing parameter between the
superpartners of the right-handed leptons, which can reach $\sim
10^{-3}$. This can enhance $R_K$ by ${\cal O}(1\%)$ without
contradicting any experimental constraints known at present,
including upper bounds on the LFV decays $\tau\to eX$ with
$X=\eta,\gamma,\mu\bar\mu$. On the other hand, $R_K$ is sensitive to
the neutrino mixing parameters within the SM extension involving a
fourth generation~\cite{la10}.

The first measurements of $R_K$ were performed in the
1970s~\cite{cl72,he75,he76}, while the current PDG world
average~\cite{pdg} is based on a more precise recent
result~\cite{am09} $R_K=(2.493\pm0.031)\times 10^{-5}$. A new
measurement of $R_K$ based on a part of a dedicated data sample
collected by the NA62 experiment at CERN in 2007 is reported here:
the analyzed $K_{e2}$ sample is $\sim 4$ times larger than the total
world sample, allowing the first measurement of $R_K$ with a
relative precision below 1\%.

The decay $K^\pm\to\pi^\mp\mu^\pm\mu^\pm$ violating lepton number by
two units can proceed via a neutrino exchange if the neutrino is a
Majorana particle, consequently the experimental limits on this
decay provide constraints on the effective Majorana neutrino mass
$\langle m_{\mu\mu}\rangle$~\cite{zu00}. This decay has also been
studied in the context of supersymmetric models with $R$-parity
violation~\cite{li00}. The best previous upper limit on the decay
rate was based on a special data set collected by the BNL E865
experiment in 1997~\cite{ap00}. The sample of $\pi\mu\mu$ triggers
collected by the NA48/2 experiment at CERN during the 2003--04 data
taking is about 8 times larger than the E865 one, which allows
improving the upper limit significantly.

\section{Beam and detector}

The NA48/2 and NA62 (phase I) experiments at CERN took data in
2003--04 and 2007--08, respectively, using the same kaon beamline
and experimental setup~\cite{fa07}. The trigger logic was optimized
to detect direct CP violating charge asymmetries in $K^\pm$ decays
in 2003--04~\cite{ba07}, and for the $K_{e2}/K_{\mu2}$ ratio
measurement in 2007--08. The beam line is capable of delivering
simultaneous unseparated $K^+$ and $K^-$ beams derived from the 400
GeV/$c$ primary proton beam extracted from the CERN SPS. Central
values of kaon momentum of 60 GeV/$c$ (both $K^+$ and $K^-$ beams)
and 74 GeV/$c$ ($K^+$ beam only), with a narrow momentum band, were
used for collection of the main data samples by the NA48/2 and NA62
experiments, correspondingly.

The fiducial decay region is contained in a 114 m long cylindrical
vacuum tank. With $\sim 10^{12}$ primary protons incident on the
target per SPS pulse of $4.8$~s duration, the typical secondary beam
flux at the entrance to the decay volume is $10^7$ to $10^8$
particles per pulse, of which about $5\%$ are kaons, while pions
constitute the dominant component. The transverse size of the beams
within the decay volume is below 1~cm (rms), and their angular
divergence is negligible.

Among the subdetectors located downstream the decay volume, a
magnetic spectrometer, a plastic scintillator hodoscope (HOD), a
liquid krypton electromagnetic calorimeter (LKr) and a muon veto
counter (MUV) are principal for the present measurements. The
spectrometer, used to detect charged products of kaon decays, is
composed of four drift chambers (DCHs) and a dipole magnet. The HOD
producing fast trigger signals consists of two planes of
strip-shaped counters. The LKr, used for particle identification and
as a veto, is an almost homogeneous ionization chamber, $27X_0$
deep, segmented transversally into 13,248 cells (2$\times$2 cm$^2$
each), and with no longitudinal segmentation. The MUV is composed of
three planes of plastic scintillator strips read out by
photomultipliers at both ends. A beam pipe traversing the centres of
the detectors allows undecayed beam particles and muons from decays
of beam pions to continue their path in vacuum.

\section{Search for lepton flavour violation}

The precision measurement of $R_K=\Gamma(K_{e2})/\Gamma(K_{\mu 2})$
is based on the NA62 2007 data sample. The measurement method is
based on counting the numbers of reconstructed $K_{e2}$ and
$K_{\mu2}$ candidates collected concurrently. Consequently the
result does not rely on kaon flux measurement, and several
systematic effects (e.g. due to reconstruction and trigger
efficiencies, time-dependent effects) cancel to first order. To take
into account the significant dependence of signal acceptance and
background level on lepton momentum, the measurement is performed
independently in bins of this observable: 10 bins covering a lepton
momentum range of $(13; 65)$~GeV/$c$ are used. The ratio $R_K$ in
each bin is computed as
\begin{equation}
R_K = \frac{1}{D}\cdot \frac{N(K_{e2})-N_{\rm
B}(K_{e2})}{N(K_{\mu2}) - N_{\rm B}(K_{\mu2})}\cdot
\frac{A(K_{\mu2})}{A(K_{e2})} \cdot
\frac{f_\mu\times\epsilon(K_{\mu2})}
{f_e\times\epsilon(K_{e2})}\cdot\frac{1}{f_\mathrm{LKr}},
\label{eq:rkcomp}
\end{equation}
where $N(K_{\ell 2})$ are the numbers of selected $K_{\ell 2}$
candidates $(\ell=e,\mu)$, $N_{\rm B}(K_{\ell 2})$ are numbers of
background events, $A(K_{\mu 2})/A(K_{e2})$ is the geometric
acceptance correction, $f_\ell$ are efficiencies of $e$/$\mu$
identification, $\epsilon(K_{\ell 2})$ are trigger efficiencies,
$f_\mathrm{LKr}$ is the global efficiency of the LKr readout, and
$D=150$ is the downscaling factor of the $K_{\mu2}$ trigger.

A detailed Monte Carlo (MC) simulation including beam line optics,
full detector geometry and material description, stray magnetic
fields, local inefficiencies of DCH wires, and time variations of
the above throughout the running period, is used to evaluate the
acceptance correction $A(K_{\mu2})/A(K_{e2})$ and the geometric
parts of the acceptances for background processes entering the
computation of $N_B(K_{\ell 2})$. Simulations are used to a limited
extent only: particle identification, trigger and readout
efficiencies are measured directly.

Due to the topological similarity of $K_{e2}$ and $K_{\mu2}$ decays,
a large part of the selection conditions is common for both decays:
(1) exactly one reconstructed positively charged particle compatible
with originating from a beam $K$ decay; (2) its momentum
$13~{\textrm{GeV}}/c<p<65~{\textrm{GeV}}/c$ (the lower limit is due
to the 10 GeV LKr energy deposit trigger requirement); (3)
extrapolated track impact points in subdetectors are within their
geometrical acceptances; (4) no LKr energy deposition clusters with
energy $E>2$~GeV not associated to the track, to suppress background
from other kaon decays; (5) distance between the charged track and
the nominal kaon beam axis ${\rm CDA}<3$~cm, and decay vertex
longitudinal position within the nominal decay volume.

The following two principal selection criteria are different for the
$K_{e2}$ and $K_{\mu2}$ decays. $K_{\ell 2}$ kinematic
identification is based on the reconstructed squared missing mass
assuming the track to be a positron or a muon:
$M_{\mathrm{miss}}^2(\ell) = (P_K - P_\ell)^2$, where $P_K$ and
$P_\ell$ ($\ell = e,\mu$) are the four-momenta of the kaon (average
beam momentum assumed) and the lepton (positron or muon mass
assumed). A selection condition
$-M_1^2<M_{\mathrm{miss}}^2(\ell)<M_2^2$ is applied to select
$K_{\ell 2}$ candidates, where $M_1^2$ varies between 0.013 and
0.016 (GeV/$c^2$)$^2$ and $M_2^2$ between 0.010 and 0.013
(GeV/$c^2$)$^2$ across the lepton momentum bins, depending on
$M_{\mathrm{miss}}^2(\ell)$ resolution. Particle identification is
based on the ratio $E/p$ of track energy deposit in the LKr
calorimeter to its momentum measured by the spectrometer. Particles
with $0.95<E/p<1.1$ ($E/p<0.85$) are identified as positrons
(muons).

\begin{figure}[tb]
\begin{center}
\resizebox{0.50\textwidth}{!}{\includegraphics{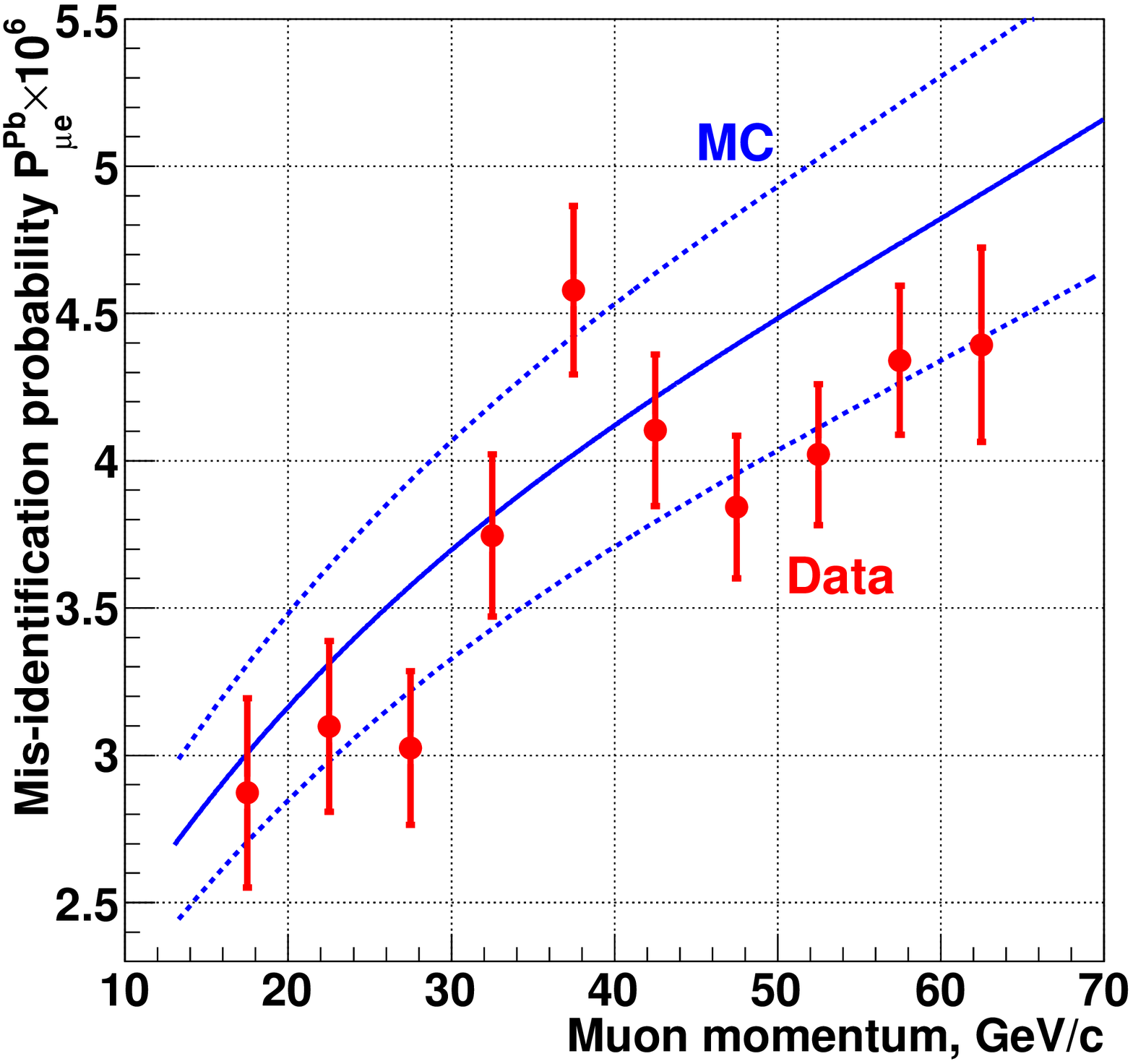}}%
\resizebox{0.50\textwidth}{!}{\includegraphics{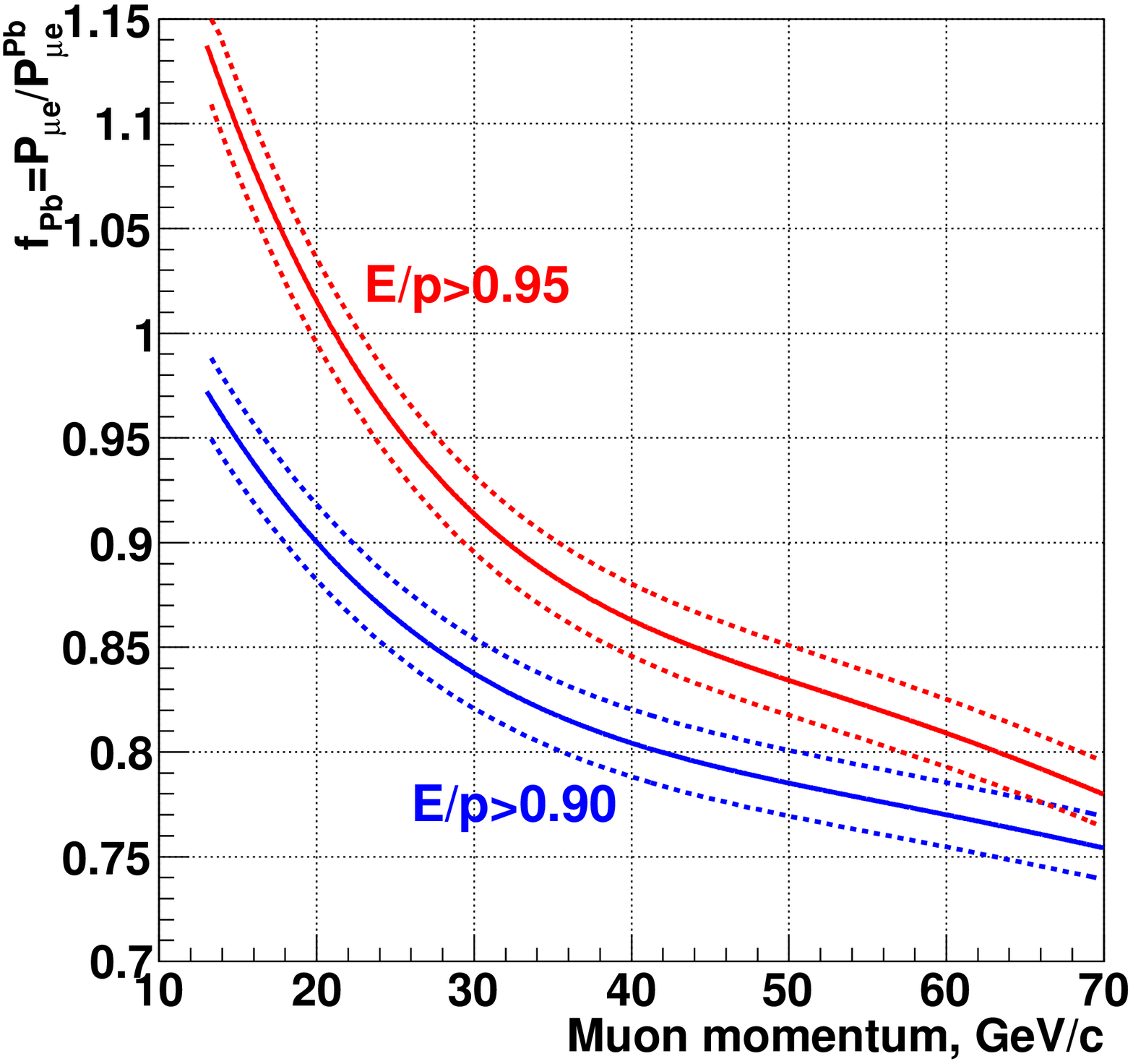}}
\end{center}
\vspace{-3mm} \caption{Left: Mis-identification probability for
muons traversing the lead wall, $P_{\mu e}^\mathrm{Pb}$, for
$(E/p)_\mathrm{min}=0.95$ as a function of momentum: measurement
(solid circles with error bars) and simulation (solid line). Right:
Correction factors $f_\mathrm{Pb}=P_{\mu e}/P_{\mu e}^\mathrm{Pb}$
for the considered values of $(E/p)_\mathrm{min}$ , as evaluated
with simulation. Dotted lines in both plots indicate the estimated
systematic uncertainties of the simulation.} \label{fig:pmue}
\end{figure}

Kinematic separation of $K_{e2}$ from $K_{\mu 2}$ decays is
achievable at low lepton momentum only ($p<35$~GeV/$c$). At high
lepton momentum, the $K_{\mu2}$ decay with a mis-identified muon
($E/p>0.95$) is the largest background source. The dominant process
leading to mis-identification of the muon as a positron is
`catastrophic' bremsstrahlung in or in front of the LKr leading to
significant energy deposit in the LKr. Mis-identification due to
accidental LKr clusters associated with the muon track is
negligible, as concluded from a study of the sidebands of
track-cluster time difference and distance distributions.

The muon mis-identification probability $P_{\mu e}$ has been
measured as a function of momentum. To collect a muon sample free
from the typical $\sim10^{-4}$ positron contamination due to $\mu\to
e$ decays, a $9.2X_0$ thick lead (Pb) wall covering $\sim 20\%$ of
the geometric acceptance was installed approximately 1.2~m in front
of the LKr calorimeter (between the two HOD planes) during a period
of data taking. The component from positrons which traverse the Pb
wall and are mis-identified as muons from $K_{\mu2}$ decay with
$p>30$~GeV/$c$ and $E/p>0.95$ is suppressed down to a negligible
level ($\sim 10^{-8}$) by energy losses in the Pb.

However, muon passage through the Pb wall affects the measured
$P_{\mu e}^\mathrm{Pb}$ via two principal effects: 1) ionization
energy loss in Pb decreases $P_{\mu e}$ and dominates at low
momentum; 2) bremsstrahlung in Pb increases $P_{\mu e}$ and
dominates at high momentum. To evaluate the correction factor
$f_\mathrm{Pb}=P_{\mu e}/P_{\mu e}^\mathrm{Pb}$, a dedicated MC
simulation based on Geant4 (version 9.2) has been developed to
describe the propagation of muons downstream from the last DCH,
involving all electromagnetic processes including muon
bremsstrahlung~\cite{ke97}.

The measurements of $P_{\mu e}^{\mathrm{Pb}}$ in momentum bins
compared with the results of the MC simulation and the correction
factors $f_\mathrm{Pb}$ obtained from simulation, along with the
estimated systematic uncertainties of the simulated values, are
shown in Fig.~\ref{fig:pmue}.  The relative systematic uncertainties
on $P_{\mu e}$ and $P_{\mu e}^{\mathrm{Pb}}$ obtained by simulation
have been estimated to be $10\%$, and are mainly due to the
simulation of cluster reconstruction and energy calibration. However
the error of the ratio $f_\mathrm{Pb}=P_{\mu e}/P_{\mu
e}^\mathrm{Pb}$ is significantly smaller ($\delta
f_\mathrm{Pb}/f_\mathrm{Pb}=2\%$) due to cancellation of the main
systematic effects. The measured $P_{\mu e}^{\mathrm{Pb}}$ is in
agreement with the simulation within their uncertainties.

The $K_{\mu 2}$ background contamination integrated over lepton
momentum has been computed to be $(6.11\pm0.22)\%$ using the
measured $P_{\mu e}^\mathrm{Pb}$ corrected by $f_\mathrm{Pb}$. The
quoted error is due to the limited size of the data sample used to
measure $P_{\mu e}^\mathrm{Pb}$ (0.16\%), the uncertainty $\delta
f_\mathrm{Pb}$ (0.12\%), and the model-dependence of the correction
for the $M_\mathrm{miss}^2(e)$ vs $E/p$ correlation (0.08\%).

$R_K$ is defined to be fully inclusive of internal bremsstrahlung
(IB) radiation~\cite{ci07}. The structure-dependent (SD) $K^+\to
e^+\nu\gamma$ process~\cite{bi93,ch08} may lead to a $K_{e2}$
signature if the positron is energetic and the photon is undetected.
In particular, the $\mathrm{SD}^+$ component with positive photon
helicity peaks at high positron momentum in the $K^+$ rest frame
($E^*_e\approx M_K/2$) and has a similar branching ratio to
$K_{e2}$. The background due to $K^+\to
e^+\nu\gamma~(\mathrm{SD}^-)$ decay with negative photon helicity
peaking at $E^*_e\approx M_K/4$ and the interference between the IB
and SD processes are negligible. The $\mathrm{SD}^+$ background
contribution has been estimated by MC simulation as
$(1.07\pm0.05)\%$, using a recent measurement of the $K^+\to
e^+\nu\gamma~(\mathrm{SD}^+)$ differential decay rate~\cite{am09}.
The quoted uncertainty is due to the limited precision on the form
factors and decay rate, and is therefore correlated between lepton
momentum bins.

\begin{figure}[tb]
\begin{center}
\resizebox{0.5\textwidth}{!}{\includegraphics{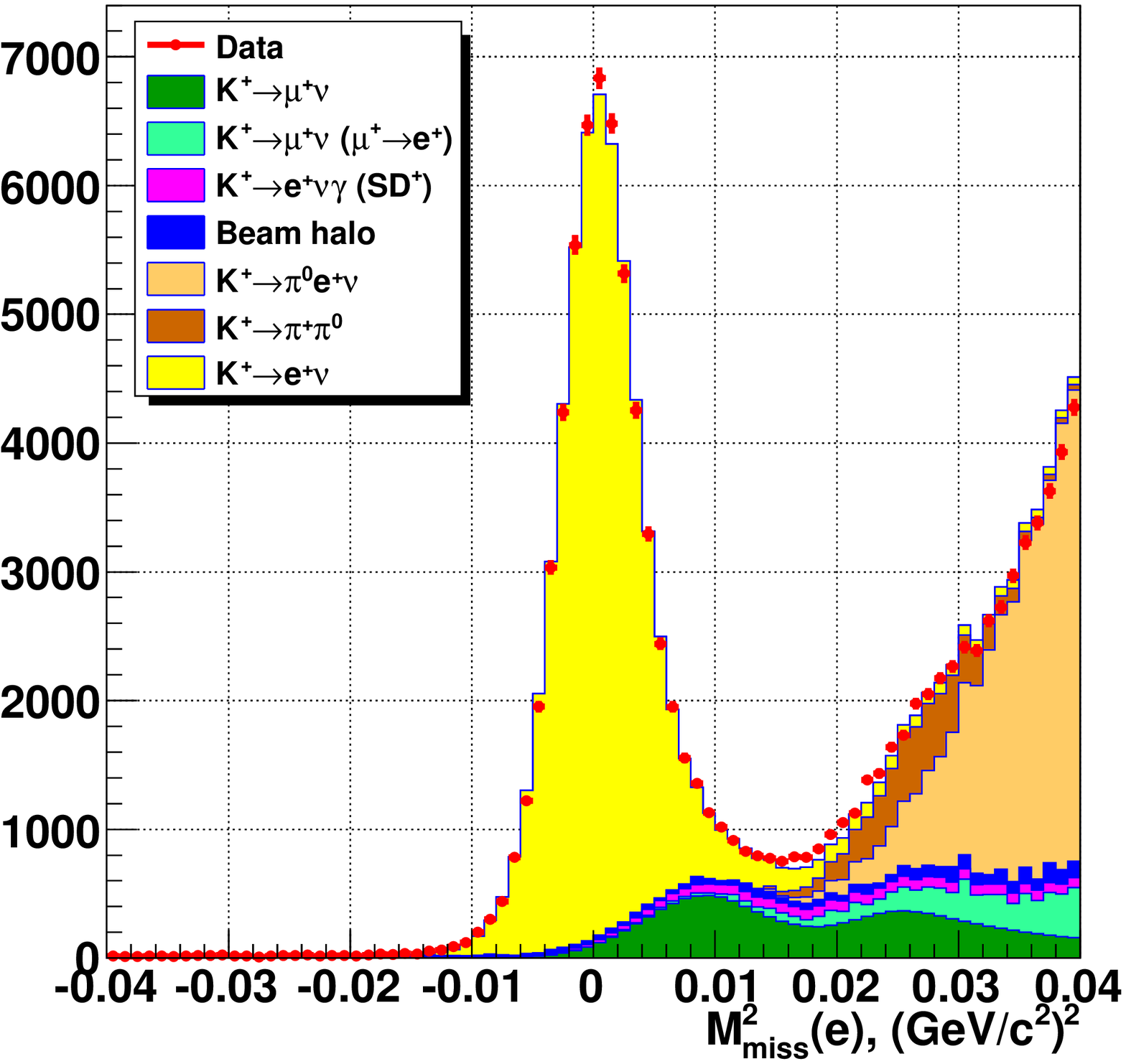}}%
\resizebox{0.5\textwidth}{!}{\includegraphics{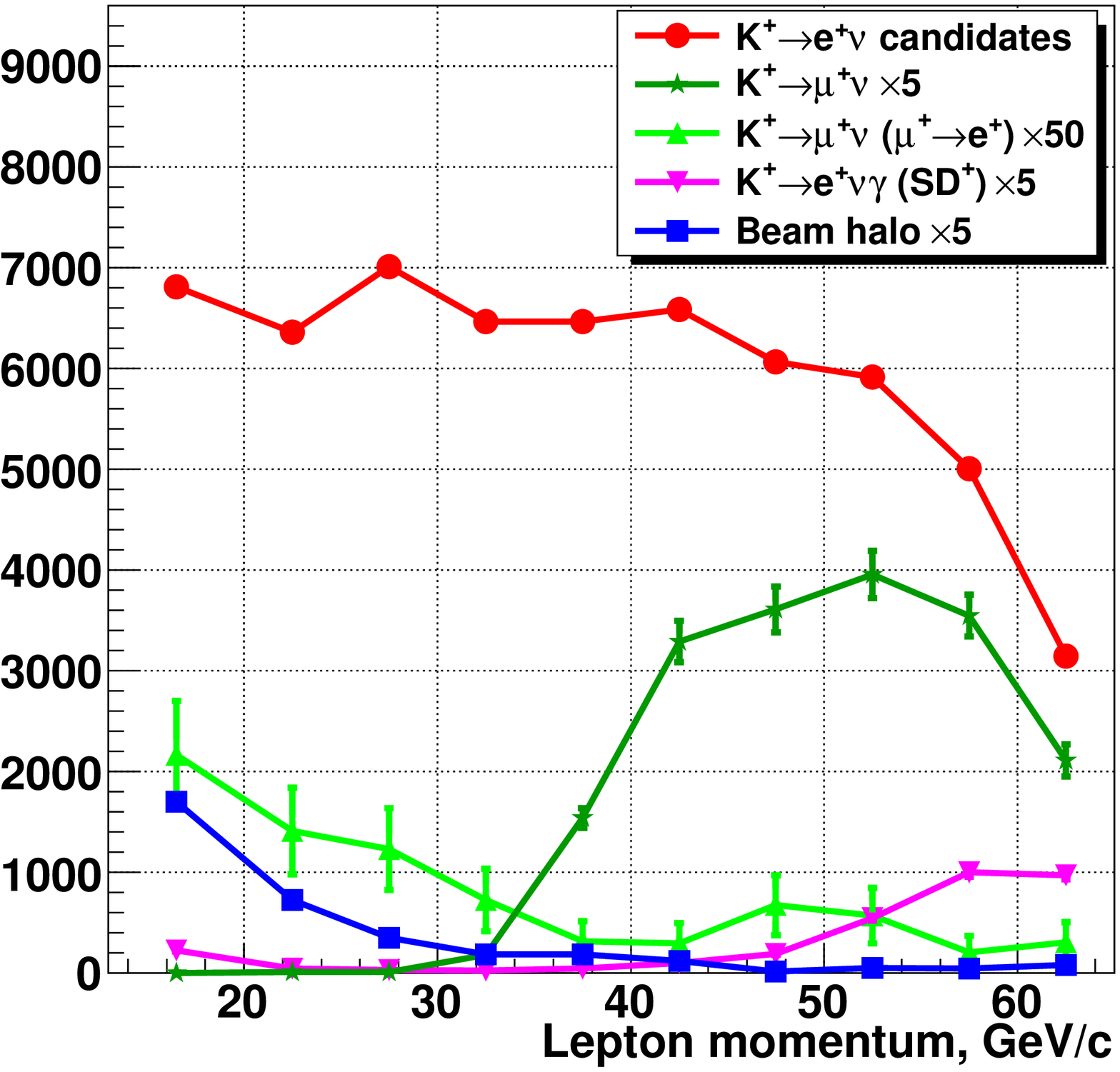}}
\end{center}
\vspace{-3mm} \caption{Left: Reconstructed squared missing mass
$M_{\mathrm{miss}}^2(e)$ distribution of the $K_{e2}$ candidates
compared with the sum of normalised estimated signal and background
components. Right: Lepton momentum distributions of the $K_{e2}$
candidates and the dominant backgrounds; the backgrounds are scaled
for visibility.} \label{fig:mm2e}
\end{figure}

\begin{table}[tb]
\begin{center}
\caption{Summary of backgrounds in the $K_{e2}$ sample.}
\label{tab:bkg} \vspace{2mm}
\begin{tabular}{|l|c|}
\hline Source                  & $N_B/N(K_{e2})$\\
\hline $K_{\mu 2}$             & $(6.11\pm0.22)\%$\\
$K_{\mu 2}(\mu\to e)$          & $(0.27\pm0.04)\%$\\
$K^+\to e^+\nu\gamma~(\mathrm{SD}^+)$ & $(1.07\pm0.05)\%$\\
$K^+\to\pi^0 e^+\nu$           & $(0.05\pm0.03)\%$\\
$K^+\to\pi^+\pi^0$             & $(0.05\pm0.03)\%$\\
Beam halo                      & $(1.16\pm0.06)\%$\\
\hline
Total                          & $(8.71\pm0.24)\%$\\
\hline
\end{tabular}
\vspace{-4mm}
\end{center}
\end{table}

The beam halo background in the $K_{e2}$ sample induced by halo
muons (undergoing $\mu\to e$ decay in flight or mis-identified) is
measured using a data-driven method, by reconstructing $K^+_{e2}$
candidates from a control $K^-$ data sample collected with the $K^+$
beam dumped, to be $(1.16\pm0.06)\%$. Background rate and
kinematical distribution are qualitatively reproduced by a halo
simulation. The uncertainty is due to the limited size of the
control sample. The beam halo is the only significant background
source in the $K_{\mu2}$ sample. Its contribution is mainly at low
muon momentum, and has been measured to be $(0.38\pm0.01)\%$ using
the same technique as for the $K_{e2}$ sample.

The numbers of selected $K_{e2}$ and $K_{\mu 2}$ candidates are
59813 and $1.803\times 10^7$, respectively (the latter samples has
been pre-scaled by a factor of 150 at the trigger level).
Backgrounds in the $K_{e2}$ sample integrated over lepton momentum
are summarised in Table~\ref{tab:bkg}: the total background
contamination is $(8.71\pm0.24)$\%, and its uncertainty is smaller
than the relative statistical uncertainty of 0.43\%. The
$M_{\mathrm{miss}}^2(e)$ and lepton momentum distributions of
$K_{e2}$ candidates and backgrounds are shown in
Fig.~\ref{fig:mm2e}.

The ratio of geometric acceptances $A(K_{\mu2})/A(K_{e2})$ in each
lepton momentum bin has been evaluated with MC simulation. The
radiative $K^+\to e^+\nu\gamma$ (IB) process, which is responsible
for the loss of about 5\% of the $K_{e2}$ acceptance by increasing
the reconstructed $M_{\rm miss}^2(e)$, is taken into account
following~\cite{bi93}, with higher order corrections according
to~\cite{we65,ga06}.

The acceptance correction is strongly influenced by bremsstrahlung
suffered by the positron in the material upstream of the
spectrometer magnet (Kevlar window, helium, DCHs). This results in
an almost momentum-independent loss of $K_{e2}$ acceptance of about
6\%, mainly by increasing the reconstructed $M_{\rm miss}^2(e)$. The
relevant material thickness has been measured by studying the
spectra and rates of bremsstrahlung photons produced by low
intensity 25~GeV/$c$ and 40~GeV/$c$ electron and positron beams
steered into the DCH acceptance, using special data samples
collected in the same setup by the NA48/2 experiment in 2004 and
2006. The material thickness during the 2007 run has been estimated
to be $(1.56\pm0.03)\% X_0$, where the quoted uncertainty is
dominated by the limited knowledge of helium purity in the
spectrometer tank.

\begin{table}
\begin{center}
\caption{Summary of the uncertainties on $R_K$.}
\label{tab:err}\vspace{2mm}
\begin{tabular}{|l|c|}
\hline
Source & $\delta R_K\times 10^5$\\
\hline
Statistical                         & 0.011  \\
\hline
~~~$K_{\mu2}$ background               & 0.005  \\
~~~$K^+\to e^+\nu\gamma~(\textrm{SD}^+)$ background & 0.001\\
~~~$K^+\to\pi^0 e^+\nu$, $K^+\to\pi^+\pi^0$ backgrounds & 0.001\\
~~~Beam halo background                & 0.001\\
~~~Helium purity                       & 0.003\\
~~~Acceptance correction               & 0.002\\
~~~Spectrometer alignment              & 0.001\\
~~~Positron identification efficiency  & 0.001\\
~~~1-track trigger efficiency          & 0.002\\
~~~LKr readout inefficiency            & 0.001\\
Total systematic                       & 0.007\\
\hline
Total & 0.013\\
\hline
\end{tabular}
\end{center}
\end{table}

\begin{figure}[tb]
\begin{center}
{\resizebox*{0.548\textwidth}{!}{\includegraphics{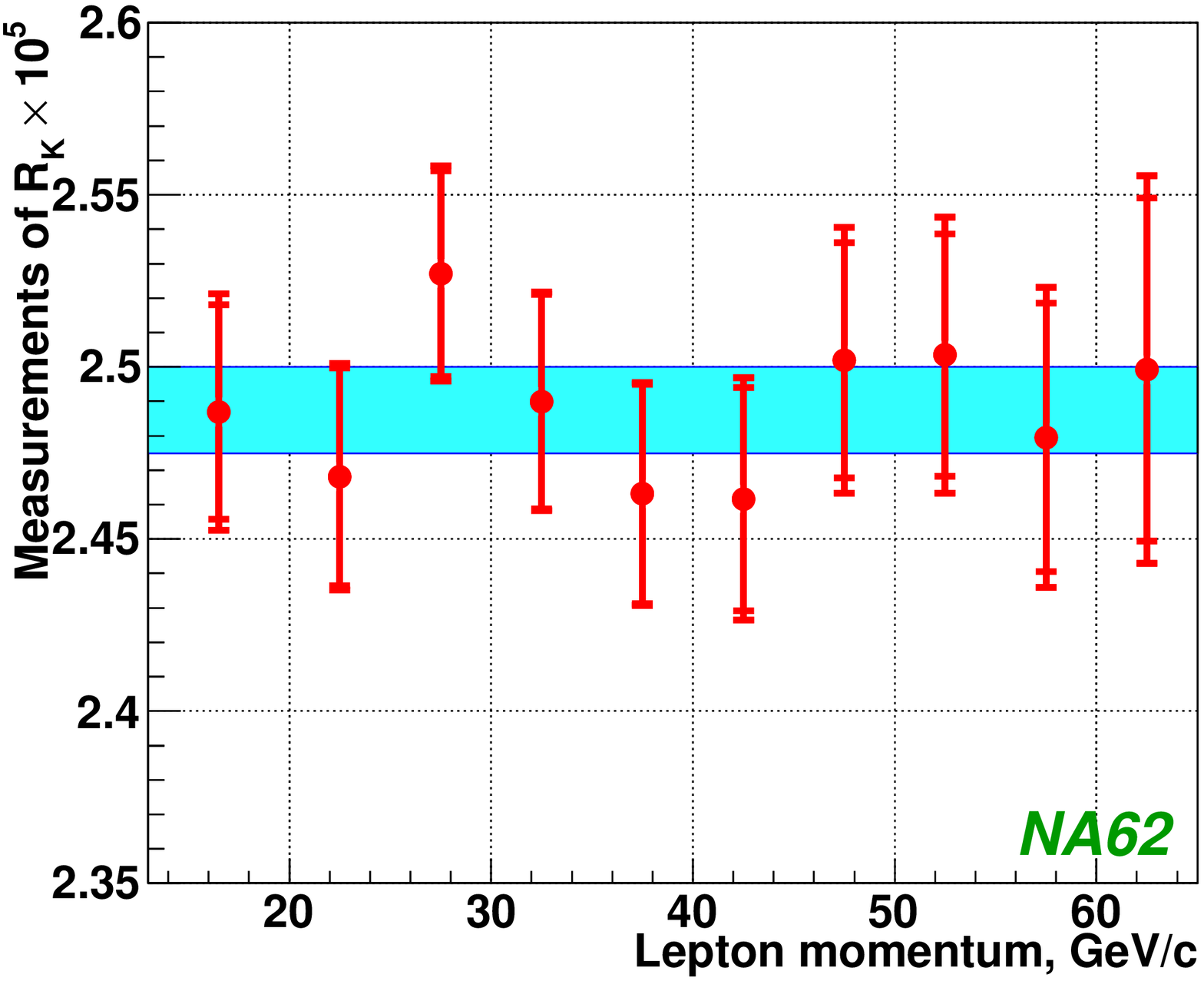}}}%
{\resizebox*{0.45\textwidth}{!}{\includegraphics{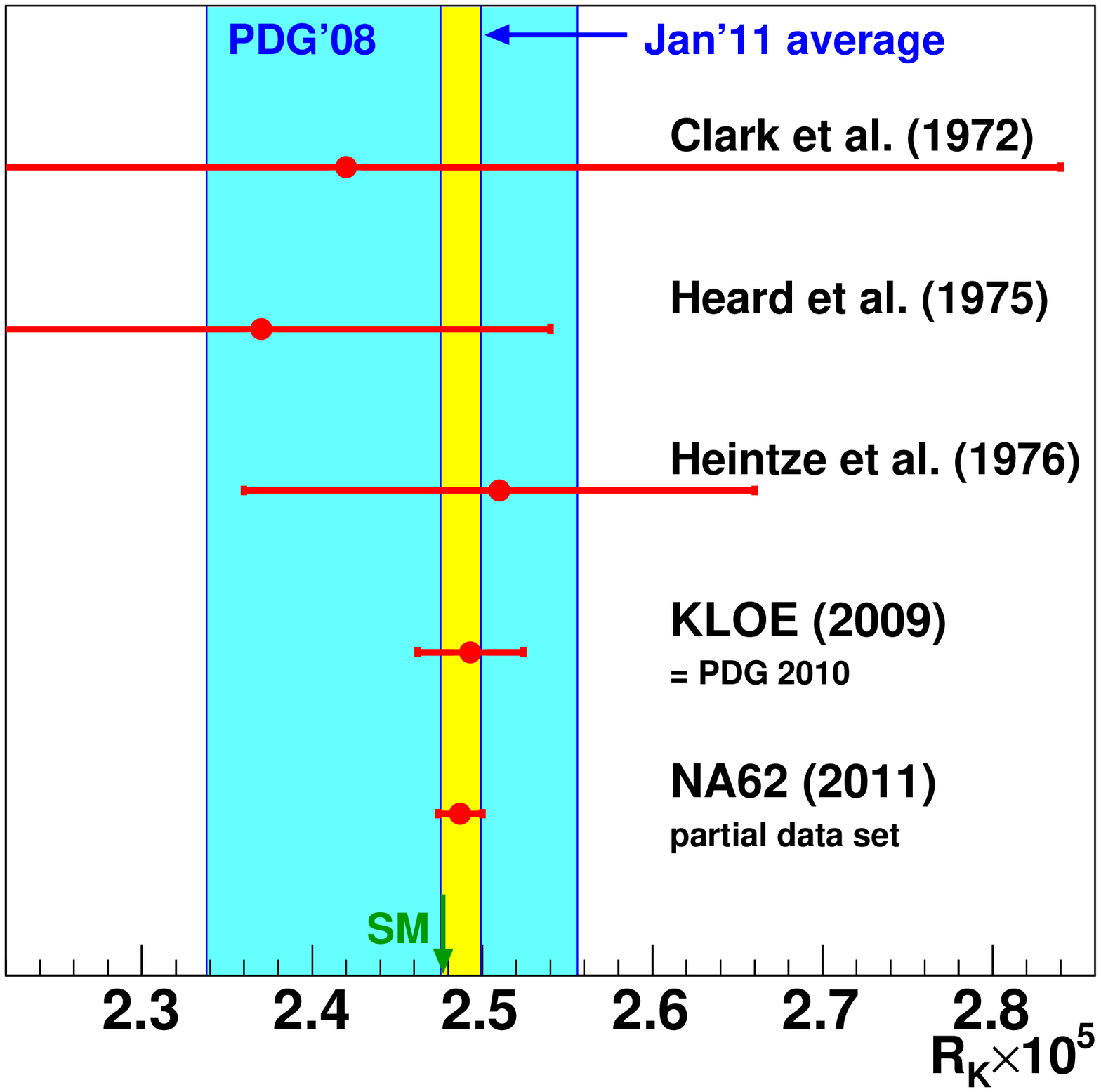}}}
\end{center}
\vspace{-3mm} \caption{Left: Measurements of $R_K$ in lepton
momentum bins with their uncorrelated statistical uncertainties and
the partially correlated total uncertainties; the average $R_K$ and
its total uncertainty are indicated by a band. Right: The new world
average including the present result.} \label{fig:rk}
\end{figure}

A $\chi^2$ fit to the measurements of $R_K$ in the 10 lepton
momentum bins has been performed, taking into account the bin-to-bin
correlations between the systematic errors. The uncertainties of the
combined result are summarized in Table~\ref{tab:err}. To validate
the assigned systematic uncertainties, extensive stability checks
have been performed in bins of kinematic variables and by varying
selection criteria and analysis procedures. The fit result
is~\cite{RK}
\begin{equation}
R_K = (2.487\pm 0.011_{\mathrm{stat.}}\pm
0.007_{\mathrm{syst.}})\times 10^{-5} =(2.487\pm0.013)\times
10^{-5},
\end{equation}
with $\chi^2/{\rm ndf}=3.6/9$. The individual measurements with
their statistical and total uncertainties, the combined NA62 result,
and the new world average are presented in Fig.~\ref{fig:rk}.

\section{Search for lepton number violation}

The $K^\pm\to\pi^\mp\mu^\pm\mu^\pm$ decay has been searched for
using the NA48/2 2003--04 data sample, normalizing to the abundant
$K^\pm\to\pi^\pm\pi^+\pi^-$ normalization channel (denoted
$K_{3\pi}$ below). Three-track vertices (compatible with either
$K^\pm\to\pi\mu\mu$ or $K_{3\pi}$ decay topology) are reconstructed
by extrapolation of track segments from the spectrometer upstream
into the decay volume, taking into account the measured Earth's
magnetic field, stray fields due to magnetization of the vacuum
tank, and multiple scattering. The vertex is required to have no
significant missing momentum, and to be composed of one $\pi^\pm$
candidate (with the ratio of energy deposition in the LKr
calorimeter to momentum measured by the spectrometer $E/p<0.95$,
which suppresses electrons, and no in-time associated hits in the
MUV), and a pair of $\mu^\pm$ candidates (with $E/p<0.2$ and
associated signal in the MUV). The muon identification efficiency
has been measured to be above $98\%$ for $p>10$~GeV/$c$, and above
$99\%$ for $p>15$~GeV/$c$.

\begin{figure}[tb]
\begin{center}
{\resizebox*{0.5\textwidth}{!}{\includegraphics{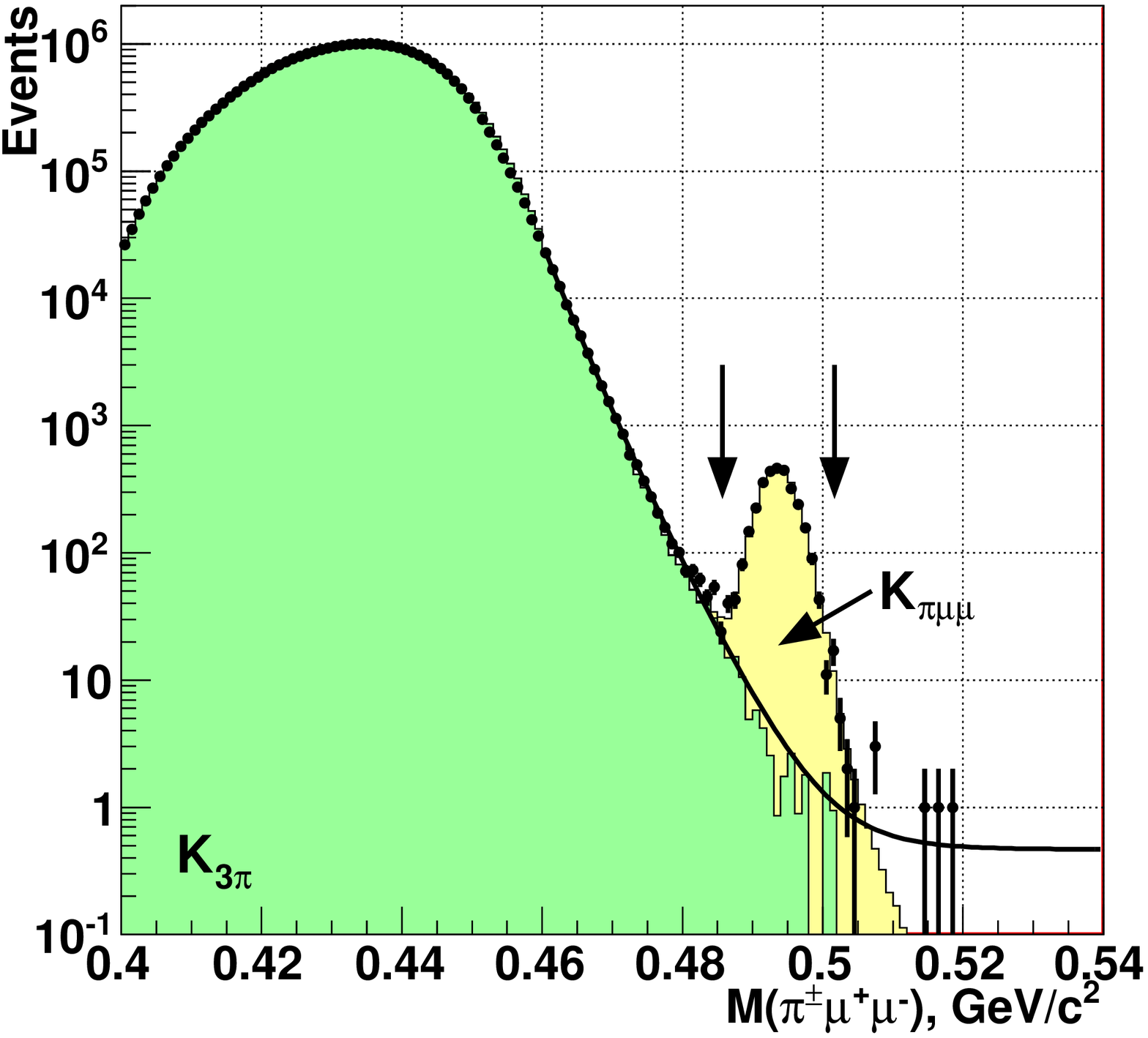}}}%
{\resizebox*{0.5\textwidth}{!}{\includegraphics{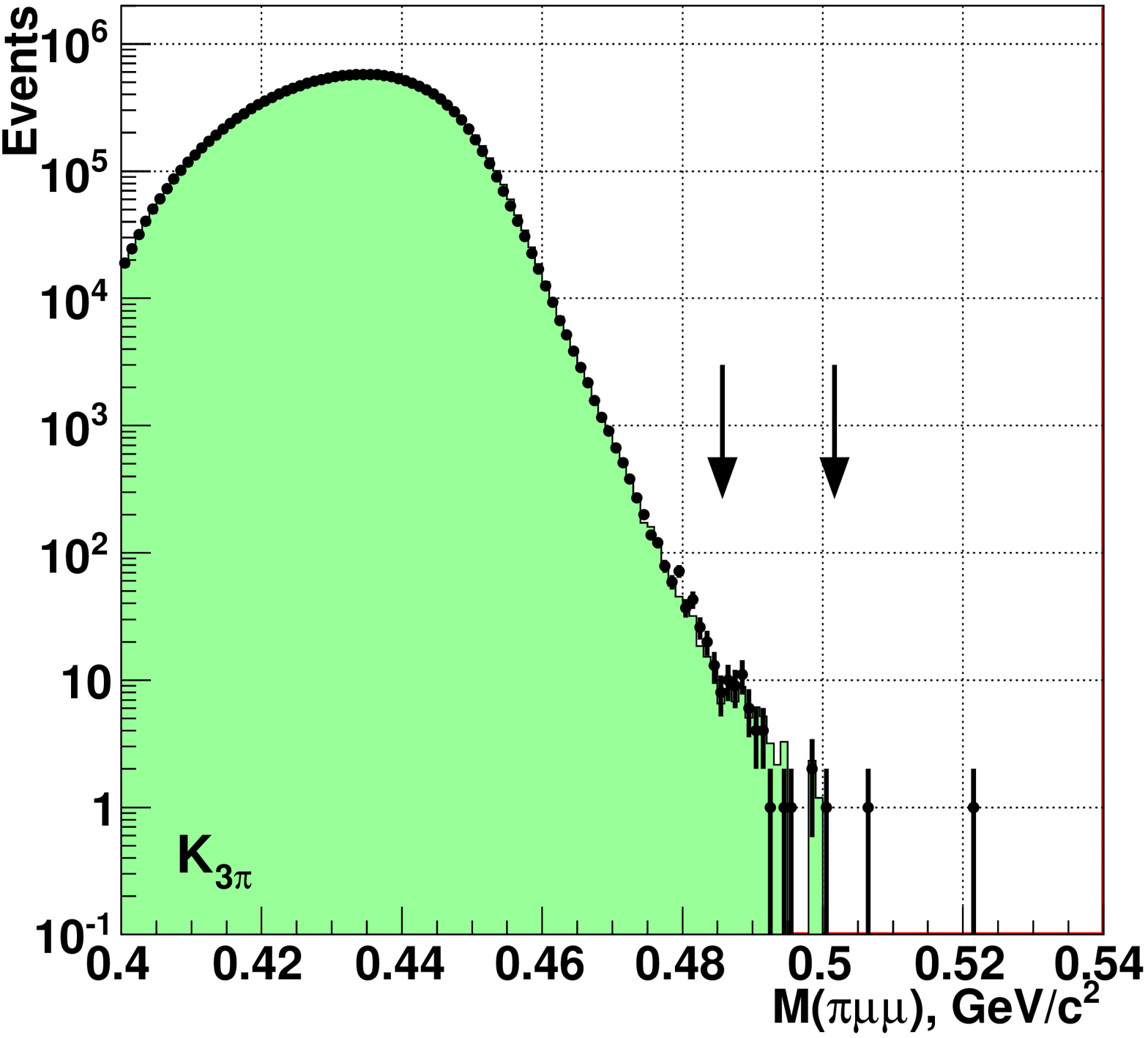}}}
\end{center}
\vspace{-3mm} \caption{Reconstructed $M_{\pi\mu\mu}$ spectra for
candidates with different (left) and same sign (right) muons: data
(dots), $K_{3\pi}$ and $K_{\pi\mu\mu}$ MC simulations (filled
areas); fit to background using the empirical parameterization as
explained in the text (solid line). The signal region is indicated
with arrows.} \label{fig:bkg}
\end{figure}

The invariant mass spectra of the reconstructed
$\pi^\pm\mu^\pm\mu^\mp$ and $\pi^\mp\mu^\pm\mu^\pm$ candidates are
presented in Fig.~\ref{fig:bkg}. The observed flavour changing
neutral current $K^\pm\to\pi^\pm\mu^\pm\mu^\mp$ decay (3120
candidates with a background of 3.3\%) has been studied
separately~\cite{ba11}. In the mass spectrum with same sign muons,
corresponding to the lepton number violating signature, 52 events
are observed in the signal region $|M_{\pi\mu\mu}-M_K|<8~{\rm
MeV}/c^2$. The background comes from the $K_{3\pi}$ decay, and has
been estimated by MC simulation to be $(52.6\pm19.8)$ events. The
quoted uncertainty is systematic due to the limited precision of MC
description of the high-mass region, and has been estimated from the
level of data/MC agreement in the control mass region of (465;
485)~MeV/$c^2$. This background estimate has been cross-checked by
fitting the mass spectrum in the region between 460 and 520~${\rm
MeV}/c^2$, excluding the signal region between 485 and 502~${\rm
MeV}/c^2$, with an empirical function similar to that used in the
E865 analysis~\cite{ap00} using the maximum likelihood estimator and
assuming a Poisson probability density in each mass bin.

The Feldman-Cousins method~\cite{fe98} is employed for confidence
interval evaluation; the systematic uncertainty of the background
estimate is taken into account. Conservatively assuming the expected
background to be $52.6-19.8=32.8$ events to take into account its
uncertainty, this translates into an upper limit of 32.2 signal
events at 90\% CL. The geometrical acceptance is conservatively
assumed to be the smallest of those averaged over the
$K^\pm\to\pi^\pm\mu^\pm\mu^\mp$ and $K_{3\pi}$ samples
($A_{\pi\mu\mu}=15.4\%$ and $A_{3\pi}=22.2\%$). This leads to an
upper limit~\cite{ba11} of ${\rm
BR}(K^\pm\to\pi^\mp\mu^\pm\mu^\pm)<1.1\times 10^{-9}$ at 90\% CL,
which improves the best previous limit~\cite{ap00} by almost a
factor of 3.

\section{Conclusions}

The most precise measurement of lepton flavour violation parameter
$R_K$ has been performed: $R_K=(2.487\pm0.013)\times 10^{-5}$ is
consistent with the SM expectation, and can be used to constrain
multi-Higgs~\cite{ma06} and fourth generation~\cite{la10} new
physics scenarios. An improved upper limit of $1.1\times 10^{-9}$
for the branching fraction of the lepton number violating
$K^\pm\to\pi^\mp\mu^\pm\mu^\pm$ decay has been established.

\end{document}